\documentclass[preprint,12pt]{elsarticle}
\usepackage{graphicx}

\usepackage{comment}
\usepackage{amssymb}
\usepackage{longtable}
\usepackage{hyperref}

\usepackage{amsthm}

\usepackage[nodots,nocompress]{numcompress}

\usepackage{bm}

\usepackage{color}

\journal{JQSRT}

\begin{document}

\begin{frontmatter}

\title{Cascade emission  in electron beam ion trap plasma of  W$^{25+}$ ion}%

\author{V. Jonauskas\corref{cor1}\fnref{label1}}
\ead{Valdas.Jonauskas@tfai.vu.lt}
\cortext[cor1]{Institute of Theoretical Physics and Astronomy, Vilnius University, 
\ A. Go\v{s}tauto 12, LT-01108  Vilnius, Lithuania}

\author[label2]{T.~P\"utterich}%
\author[label1]{S. Ku\v{c}as}
\author[label1]{\v{S}. Masys}
\author[label1]{A. Kynien\.{e}}
\author[label1]{G. Gaigalas}
\author[label1]{R. Kisielius}
\author[label1]{{L}.~Rad\v{z}i\={u}t\.{e}}
\author[label1]{P. Rynkun}
\author[label1]{G. Merkelis}

\address[label1]{Institute of Theoretical Physics and Astronomy, Vilnius University, 
\ A. Go\v{s}tauto 12, LT-01108  Vilnius, Lithuania}

\address[label2]{Max--Planck--Institut f\"ur Plasmaphysik, EURATOM Association, D--85748 Garching, Germany}

\begin{abstract}

Spectra of the W$^{25+}$ ion are studied using the collisional-radiative model
(CRM) with an ensuing cascade emission. It is determined that the cascade 
emission boosts intensities only of a few lines in the $10 - 30$ nm range. 
The cascade emission is responsible for the disappearance of structure of lines 
at about 6 nm in the electron beam ion trap plasma.  Emission band at 4.5 to 
5.3 nm is also affected by the cascade emission. The strongest lines in the CRM 
spectrum correspond to $4d^{9} 4f^{4} \rightarrow 4f^{3}$ transitions, while 
$4f^{2} 5d \rightarrow 4f^{3}$ transitions arise after the cascade emission is 
taken into account. 

\end{abstract}


\end{frontmatter}

\newpage

\section{Introduction}

Tungsten emission has been intensively studied over the last few decades due to 
its application in fusion devices \cite{Bolt_2002jnm_307_43}. The intense lines 
in tungsten spectra are observed at around 5 nm (mostly of N-shell ions) where a 
large number of transitions from several charge states contributes to the plasma 
emission. This region has attracted great attention due to its importance for 
the plasma power balance and possible diagnostic applications. The $10 - 30$ nm 
region has been also investigated in the fusion and electron beam ion trap 
(EBIT) device plasma \cite{2005jpb_38_3071_putterich, 2007apmidf_13_45_radtke, 
2008ppcp_50_085016_Putterich, 2009adas_biedermann, 2011jpb_44_175004_suzuki, 
2013acp_1545_132_putterich}. In the fusion plasma, a complex structure of lines 
is observed in this region. Surprisingly, the EBIT spectra corresponding to the 
W$^{15+}$ -- W$^{28+}$ ions feature only a few lines in the $13-18$ nm 
wavelength range \cite{2009adas_biedermann}. The emission from many tungsten 
ions contributes to the line of sight measurements in the fusion plasma, thus, 
such spectra contain many lines from the  different ions. On the other hand, the 
EBIT devices provide an unique opportunity to study the emission from one or 
several neighboring  ions. Therefore,  analysis of their spectra is much easier 
compared with those from  other plasma sources. The emission  originating only 
from  several ions can be the reason why the EBIT spectra are sparse of the 
lines in the $13 - 18$ nm range. However, the corona modeling of the spectral 
line intensities provides complex structure of the lines for  W$^{25+}$ 
\cite{2014jqsrt_136_108_alkauskas} in the aforementioned range. Since these 
calculations contradict the EBIT observations, it is necessary to check what 
kind of spectra corresponds to the collisional-radiative modeling (CRM).
On the other hand, it is shown that the cascade emission boosts the intensities 
only for some lines of the W$^{13+}$ ion in the EBIT plasma 
\cite{2013jqsrt_127_64_jonauskas}.  It has to be noted that the term 
``cascade emission'' is used here instead of the radiative cascade in order to 
distinguish population of the levels from the higher-lying levels through the 
radiative cascade which is accounted in the corona model. The different
 population mechanisms appear on the scene in these two cases 
\cite{2013jqsrt_127_64_jonauskas}. The cascade processes were mostly studied 
for the radiative and Auger decays when an  inner-shell vacancy  was created 
\cite{2003jpb_36_4403_jonauskas, 2008jpb_41_215005_jonauskas, 
2010pra_82_043419_palaudoux, 2011pra_84_053415_jonauskas}.  
 
Ions in the EBIT move in cycloidal orbits spending part of their time outside 
electron beam \cite{1995ps_59_392_gillaspy}. The cascade emission starts when 
the ions leave the electron beam and the interaction with the electrons ends. 
This effect is more pronounced for the ions in the low or intermediate 
ionization stages \cite{2009apj_702_838_liang}. It was found that under the same 
conditions, the higher charge ions show less expansion in the radial direction. 
When the charge state of the ions increases, the Coulomb's attraction force 
directed toward the electron beam also increases. It leads to the decrease of 
the time the ions spend outside the beam where the cascade emission depopulates 
the excited levels. The effective electron density is often introduced in order 
to reduce electron-impact collision rates \cite{2009apj_702_838_liang, 
2004apj_611_598_chen}. The time fraction which the ions spend inside and outside 
the electron beam depends on many parameters, such as ion temperature, electron 
beam energy, electron beam current, electric and magnetic fields. On the other 
hand, the range of the ion radius $r_{i}$ ratio against the geometric electron 
radius $r_{e}$ can be expressed through the effective charge $Z_{eff}$ of the 
ion: $1/(Z_{eff}/Z)^{\alpha}$ with $1 < \alpha \leqslant 2$ 
\cite{2009apj_702_838_liang}. For  W$^{25+}$, one can estimate that the main 
paths of the ions span outside the electron beam: 
$r_{i}/r_{e} \approx 3^{\alpha}$.

The main aim of the current work is to study the emission spectra of the 
W$^{25+}$ ion in the EBIT plasma by performing the CRM with ensuing  cascade 
emission. The emission from  W$^{25+}$ has not deserved wide attention so far 
since calculations are complicated due to the open $f$ shells. Systems with the 
open $f$ shells is further of interest for the study of the complex 
multi-electron high-Z ions. The present work focuses on the $2 - 30$ nm region 
which accumulates the main emission from the W$^{25+}$ ion 
\cite{2014jqsrt_136_108_alkauskas}. As far as we know, influence of the cascade 
emission on the formation of lines in the EBIT plasma has not been studied for 
this ion before. Previous works concentrated on analysis of the spectral lines 
obtained from the CRM or the corona model \cite{2007apmidf_13_45_radtke, 
2014jqsrt_136_108_alkauskas}. It was shown that the corona model is suitable for 
the low density EBIT plasma \cite{Jonauskas2007jpb_40_2179}. The CRM 
calculations included the $4f^{3}$ and $4f^{2} 5l$ ($l=0,1,2,3$) configurations 
but omitted the important the $4d^{9} 4f^{4}$ and $4f^{2} 5g$ configurations 
\cite{2007apmidf_13_45_radtke}. Only the strongest lines were presented in their 
work. The current study has been extended to 19612 levels compared with 13937 
levels used in the corona model calculations \cite{2014jqsrt_136_108_alkauskas}.

The rest of the paper is organized as follows. In the next section we present 
theoretical methods used to calculate atomic data and emission spectra. 
In Section 3, the determined emission spectra corresponding to the CRM and the 
cascade emission are discussed.

\section{Theoretical methods}

Energy levels, radiative transition probabilities, and electron-impact 
excitation rates for W$^{25+}$ have been calculated using Flexible Atomic Code 
(FAC) \cite{2008cjp_86_675_Gu} which implements the relativistic 
Dirac-Fock-Slater method. Previous study included 22 configurations 
\cite{2014jqsrt_136_108_alkauskas} while the current work employs 43 
configurations: $4f^{3}$, $4f^{2} 5l$ ($l=0,1,2,3,4$), $4f^{2} 6l'$, 
$4f^{2} 7l'$ ($l'=0,1,2,3,4,5$), $4f^{2} 8l$, $4d^{9}4f^{4}$, 
$4d^{9}4f^{3}5l''$ ($l''=0,1,2$), $4d^{9}4f^{2}5s^{2}$, $4d^{8}4f^{5}$, 
$4f5s^{2}$, $4f5s5l'''$ ($l'''=1,2,3,4$), $4f5p^{2}$, $4f5p5d $, $4f5s6l$,  
$4p^{5} 4f^{4}$, $4p^{5}4f^{3}5s$. These configurations produce 19612 levels. 
Configuration interaction has been taken into account for all the considered 
configurations. The radiative transition probabilities have been calculated 
for electric dipole, quadrupole, and octupole and for magnetic dipole and 
quadrupole transitions. 

Electron-impact excitation cross-sections are obtained within the distorted wave 
approximation. Collision rates are calculated for 790 eV electron beam energy 
which corresponds to the energy used in the spectra measurements 
\cite{2009adas_biedermann}. The Gaussian distribution function with a full width 
at a half-maximum of 30 eV is used for the electron energy.

Populations of levels in the CRM have been obtained by solving the system of 
coupled rate equations
\begin{equation}
\frac{d n_{i}(t)}{dt} = N_{e} \sum_{k} n_{k}(t) C_{ki} +  
\sum_{k>i} n_{k}(t) A^{r}_{ki} - N_{e} n_{i}(t) 
\sum_{k} C_{ik} - n_{i}(t) \sum_{j<i} A^{r}_{ij}
\label{crm}
\end{equation} 
in the steady-state equilibrium approximation ($\frac{d n_{i}}{dt} = 0$). 
Here $n_{i}$ is the population of the level $i$, $A_{ij}^{r}$ is the  radiative 
transition probability from the level $i$ to the level $j$, and $C_{ik}$ is the 
electron-impact excitation rate from the level $i$ to the level $k$, $N_{e}$ is 
the electron density ($N_{e} = 1 \times 10^{12}$ cm$^{-3}$). 

Total populations of the levels during the cascade emission can be found by 
summation of the population in every step of the cascade:
\begin{equation}
n_{i}^{j+1}=\sum_{m>i} \frac{n_{m}^{j} A_{mi}^{r}}{\sum_{k<m} A_{mk}^{r}}, 
\label{radcas}
\end{equation} 
where $n_{i}^{j}$ corresponds to the population of the level $i$ in $j$ step of 
the cascade. By the step of the cascade, we mean all possible radiative 
transitions from every not zero-populated level to the other levels. Thus, 
transfer of the population through the intermediate levels is not included in 
the single step. Equation (\ref{radcas}) means that radiative transition from 
the level $m$ to the level $i$ transfers only part 
${A_{mi}^{r}}/{\sum_{k<m} A_{mk}^{r}}$ of the population $n_{m}^{j}$. 
The same approach was used analyzing Auger cascades 
\cite{2003jpb_36_4403_jonauskas, 2010pra_82_043419_palaudoux, 
2011pra_84_053415_jonauskas}. Since the cascade emission takes place after ions 
leave electron beam, the initial population of the levels for the first step of 
the cascade is determined from the CRM.

Equation (\ref{radcas}) determines the populations of the levels when all the 
higher-lying levels are depopulated by the radiative decay. However, fraction of 
depopulation strongly depends on the time which the ions spend outside the 
electron beam. In this case, the population of the levels has to be determined 
by solving the time-dependent rate equations which omits interaction with the 
electrons:
\begin{equation}
\frac{d n_{i}(t)}{dt} = \sum_{k>i} n_{k}(t) A^{r}_{ki} - n_{i}(t) \sum_{j<i} A^{r}_{ij}.
\label{tdradcas}
\end{equation} 

The total population $n_{i}(\Delta t)$ which leaves the level $i$ during the 
time interval $\Delta t$ is found by integrating expression:
\begin{equation}
\frac{d n_{i}(t)}{dt} = n_{i}(t) \sum_{j<i} A_{ij}^{r}
\label{timepopulation}
\end{equation} 
which leads to 
\begin{equation}
n_{i}(\Delta t) = \int_{0}^{\Delta t} \frac{d n_{i}(t)}{dt} dt = \int_{0}^{\Delta t} 
n_{i}(t) dt \sum_{j<i} A_{ij}^{r}.
\label{population}
\end{equation} 
Here $\Delta t$ is the time the ions have spent outside the electron beam. 
Equation (\ref{population}) provides the populations for the time-integrated 
line intensities. The total population obtained by summation of the population 
from every step of the cascade in  Eq. (\ref{radcas}) corresponds to the 
integration taking $\Delta t = \infty$ in Eq. (\ref{population}). Practically, 
however, convergence of the spectral line intensities has to be obtained for the 
finite time values. The equation (\ref{radcas}) is applied to calculate the 
final spectra of the cascade emission. Thus, there is no need to perform 
convergence check of the spectra obtained from Eq. (\ref{population}).

\section{Results}

\begin{figure}
 \includegraphics[scale=0.5]{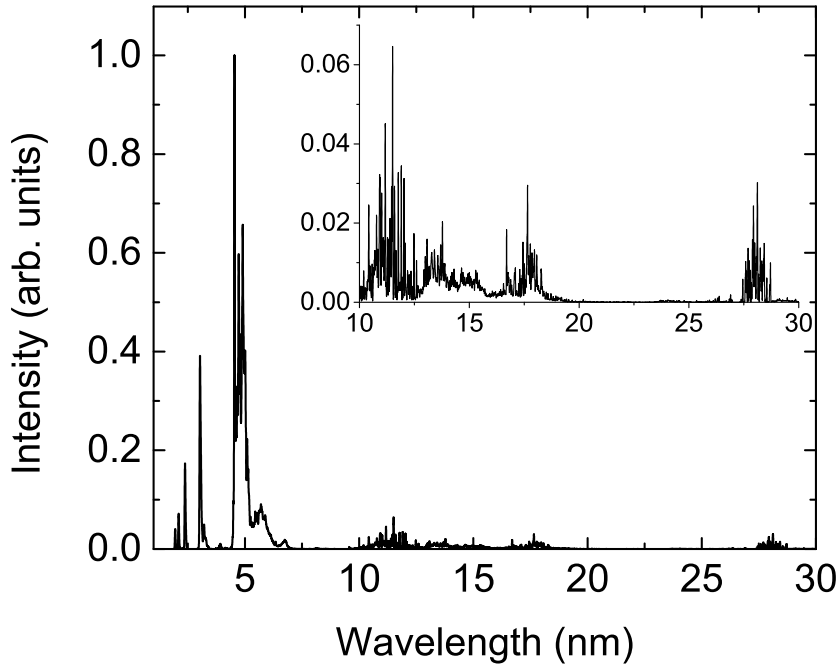}
 \caption{\label{fig1} CRM spectrum of  W$^{25+}$. The inset shows structure of 
lines in the range $10 - 30$ nm.}
\end{figure}

The CRM spectrum in the $2-30$ nm range is presented in Fig. \ref{fig1}. 
The strongest lines correspond to the $4d^{9}4f^{4} \rightarrow 4f^{3}$ and 
$4f^{2}5d \rightarrow 4f^{3}$ transitions. These lines form complex structure at 
about 5 nm. It has to be noted that our CRM calculations succeeded to reproduce 
a smaller peak in the $5.5 - 6$ nm  region. The similar peak but with larger 
intensity was obtained for  W$^{23+}$ in the CRM spectra at various electron 
densities using Maxwellian distribution for the electron velocities 
\cite{2013acp_1545_132_putterich}. The spectra from fusion plasma contain this 
additional structure of the lines \cite{2005jpb_38_3071_putterich}. 
The following lines mainly arise from the $4d^{9} 4f^{4} \rightarrow 4f^{3}$ 
transitions in our calculations. However, this structure is not seen in the 
EBIT plasma of tungsten ions \cite{Radtke2001PhysRevA_64_012720}  suggesting 
that some other mechanisms are responsible for the line formation. 

The current CRM calculations (Fig. \ref{fig1}) and the previous results from 
the corona model \cite{2013jqsrt_127_64_jonauskas} present a complex structure 
of the emission lines  in the range $10 - 30$ nm.  In the current calculations, 
the number of configurations has been increased to check the influence of 
higher-lying levels on the formation of spectral lines. It has to be noted that 
the strongest lines in the spectral range arise from the 
$4f^{2} 5s \rightarrow 4f^{3}$ transitions which have wavelengths in the 
$10 - 12$ nm region. The configuration $4f^{2} 5s$ is the first excited one 
which can decay to the ground configuration only through the electric octupole 
transitions in a single-configuration approximation. Extended basis of 
interacting configurations makes it possible for the electric dipole transitions 
to occur. However, their transition probabilities are much lower than those of 
other electric dipole transitions in the region.  Other strong lines in this 
region come from  the $4f^{2} 5f \rightarrow 4f^{2} 5d$ ($12 - 16$ nm),  
$4f^{2} 5d \rightarrow 4f^{2} 5p$ ($12 - 14$, $16 - 18$ nm), and 
$4f^{2} 5p \rightarrow 4f^{2} 5s$ ($16 - 19$, $27 - 30$ nm) transitions.

Unfortunately, the EBIT spectra exhibit just a few lines in the spectral range 
from 13 to 18 nm \cite{2009adas_biedermann}. As the theoretical spectra contain 
the complex structure of lines compared with the observations, it was suggested  
that cascade emission process, which starts after ions leave the electron beam, 
could be important in the formation of the spectral lines. It has been 
previously demonstrated  that the cascade emission  highlights only a few lines 
in the spectrum \cite{2013jqsrt_127_64_jonauskas}. However, such an effect has 
been determined for the low ionization stage, W$^{13+}$. As was mentioned above, 
influence of the cascade emission has to be larger for the lower ionization 
stages compared to the intermediate charge states, which, as far as we know, 
have never been studied using the cascade emission process.

\begin{figure}
 \includegraphics[scale=0.47]{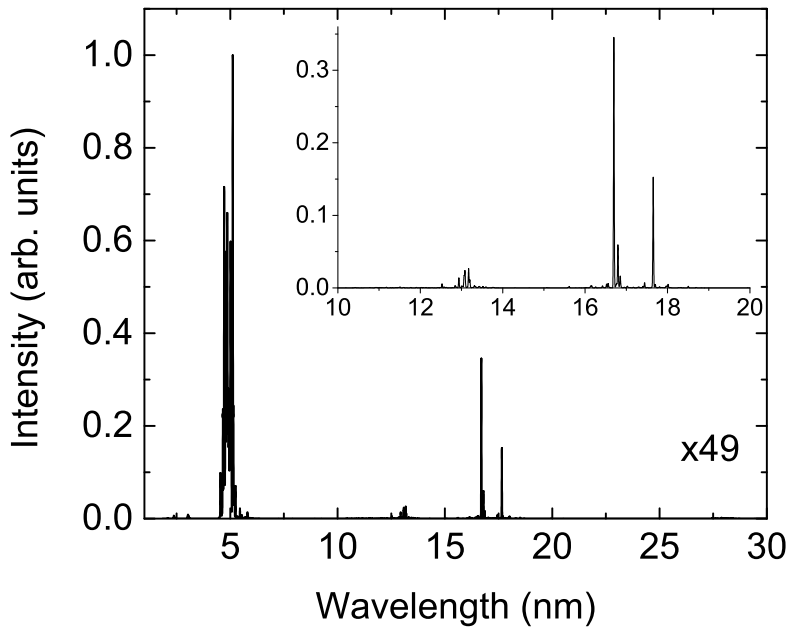}
 \caption{\label{fig2} Cascade emission spectrum of  W$^{25+}$. The inset shows 
structure of lines in the $10 - 20$ nm range. The factor shows an 
increase of the line intensities compared to the CRM spectrum.}
\end{figure}

Figure \ref{fig2} shows that the cascade emission highlights several lines in 
the range $10 - 30$ nm for the W$^{25+}$ ion. In this case, the population of 
levels is  obtained using  Eq. (\ref{radcas}). In our view, the presented 
results demonstrate the validity of our idea that the cascade emission is 
responsible for line formation in the EBIT spectra.  The strongest lines 
correspond  to the $4f^{2} 5d \rightarrow 4f^{2} 5p$ and 
$4f^{2} 5p \rightarrow 4f^{2} 5s$ transitions in  W$^{25+}$ among the levels 
with high $J$ values (Table \ref{t1}). Due to the the significantly smaller 
number of such levels and selection rules for the electric dipole transitions, 
the cascade emission leads to the concentration of intensity.  


\begingroup 

\renewcommand{\arraystretch}{1.2}
\renewcommand{\tabcolsep}{2.5mm}
\scriptsize

\begin{flushleft}
\LTcapwidth 13cm

\begin{longtable}{ l l l l l l l l  }

\caption{\label{t1} The strongest lines of the cascade emission spectrum for 
$W^{25+}$ in the $10-20$ nm wavelength range. Wavelengths $\lambda$, relative 
intensities $I$, and indexes of initial $i$ and final $f$ levels are presented. 
$J$ stands for the total angular momentum quantum number. } \\

\hline
$\lambda$ (nm) & $I$ & $i$ & $f$ & $J_{i}$ & $J_{f}$ & Initial level & Final level \\
\hline
\endfirsthead
\caption[]{ (continued) }  \\
\hline
$\lambda$ (nm) & $I$ & $i$ & $f$ & $J_{i}$ & $J_{f}$ & Initial level & Final level \\
\hline
\endhead
\hline \multicolumn{4}{r}{\textit{Continued on next page}} \\
\endfoot
\hline
\endlastfoot
$16.698$ &$100$ & $1094$ & $107$ & $17/2$ & $15/2$ & $4f_{7/2}^{2}$  $(6)$  $5d_{5/2}^{1}$   & $4f_{7/2}^{2}$  $(6)$  $5p_{3/2}^{1}$  \\
$17.653$ & $54$ & $107$ & $51$ & $15/2$ & $13/2$ & $4f_{7/2}^{2}$  $(6)$  $5p_{3/2}^{1}$   & $4f_{7/2}^{2}$  $(6)$  $5s_{1/2}^{1}$  \\
$16.693$ & $25$ & $1152$ & $127$ & $17/2$ & $15/2$ & $4f_{5/2}^{1}$  $4f_{7/2}^{1}$  $(6)$  $5d_{5/2}^{1}$   & $4f_{5/2}^{1}$   $4f_{7/2}^{1}$  $(6)$  $5p_{3/2}^{1}$  \\
$16.796$ & $21$ & $1061$ & $99$ & $15/2$ & $13/2$ & $4f_{5/2}^{1}$   $4f_{7/2}^{1}$  $(5)$  $5d_{5/2}^{1}$  & $4f_{5/2}^{1}$   $4f_{7/2}^{1}$  $(5)$  $5p_{3/2}^{1}$  \\
$13.173$ & $9$ & $1029$ & $76$ & $15/2$ &	$13/2$ & $4f_{7/2}^{2}$  $(6)$  $5d_{3/2}^{1}$  $(3)$ & $4f_{7/2}^{2}$  $(6)$  $5p_{1/2}^{1}$  \\
$13.085$ & $8$ & $1003$ & $69$ & $13/2$ & $11/2$ & $4f_{5/2}^{1}$   $4f_{7/2}^{1}$  $(5)$  $5d_{3/2}^{1}$   & $4f_{5/2}^{1}$    $4f_{7/2}^{1}$   $(5)$  $5p_{1/2}^{1}$ \\
$16.852$ & $6$ & $1087$ & $109$ & $15/2$ & $13/2$ & $4f_{7/2}^{2}$  $(6)$  $5d_{5/2}^{1}$   & $4f_{7/2}^{2}$  $(6)$  $5p_{3/2}^{1}$  \\
$13.062$ & $5$ & $1038$ & $76$ & $15/2$ & $13/2$ & $4f_{7/2}^{2}$  $(6)$  $5d_{3/2}^{1}$   & $4f_{7/2}^{2}$  $(6)$  $5p_{1/2}^{1}$  \\
$12.937$ & $5$ & $1109$ & $85$ & $15/2$ & $13/2$ & $4f_{7/2}^{2}$  $(6)$  $5d_{3/2}^{1}$   & $4f_{5/2}^{1}$   $4f_{7/2}^{1}$  $(6)$  $5p_{1/2}^{1}$  \\
\end{longtable}
\end{flushleft}
\endgroup

The EBIT experiment for the W$^{25+}$ ion also contains few strong lines but 
with shorter wavelengths  by about 2 nm compared with our calculations. 
The discrepancy between the theoretical and experimental wavelengths shows that 
important correlation effects are not taken into account for the 
$4f^{2} 5d \rightarrow 4f^{2} 5p$ and $4f^{2} 5p \rightarrow 4f^{2} 5s$ 
transitions. The importance of the correlation effects for tungsten ions has 
been illustrated for magnetic dipole transitions 
\cite{2010pra_81_012506_jonauskas, 2012adndt_98_19_jonauskas} using 
configuration interaction strength \cite{Karazija1, 1997ps_55_667_kucas} 
to build basis of the interacting configurations. However, these calculations 
are very cumbersome. Furthermore, for Er-like tungsten it was found that FAC can 
show discrepancy for wavelengths within 2 nm of the measured values when the 
correlation effects are not considered \cite{2010jpb_43_144009_clementson}.

\begin{figure}
 \includegraphics[scale=0.43]{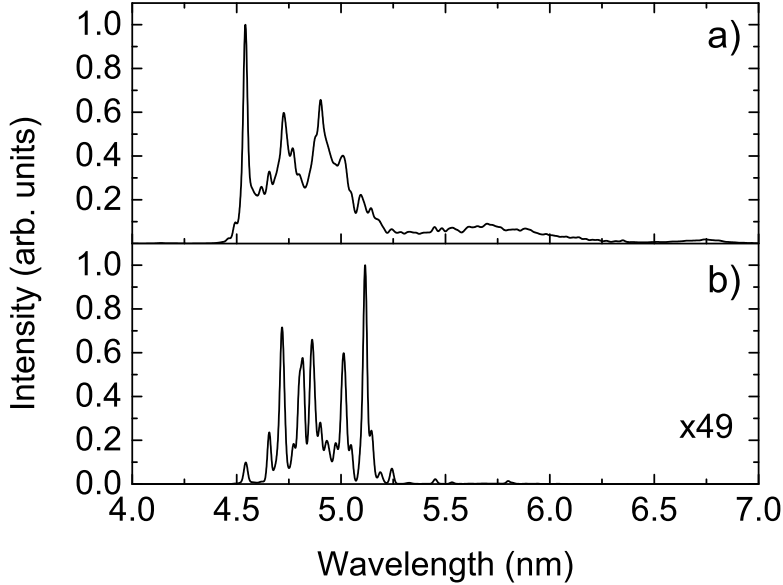}
 \caption{\label{fig3} Theoretical spectra of the W$^{25+}$ ion obtained 
a) from the CRM and b) from the CRM with ensuing emission cascade 
[Eq. (\ref{radcas})] in the $4 - 7$ nm spectral range. The factor shows an 
increase of the line intensities compared to the CRM spectrum.}
\end{figure}

It has to be noted that influence of the $4f^{2} 5s \rightarrow 4f^{3}$ 
transitions on the line formation is negligible in the cascade emission 
spectrum. The corona model has revealed that intensities of these lines 
strongly increase due to contributions of the higher-lying levels through 
radiative cascade \cite{2014jqsrt_136_108_alkauskas}.

The relative intensities of the lines in the $4 - 7$ nm region compared with the 
lines at the shorter wavelength side are strongly increased in the cascade 
emission spectrum (Fig. \ref{fig2}) compared with the CRM spectrum 
(Fig. \ref{fig1}). These lines in the $2 - 4$ nm range originate from the 
$4f^{2} 5g \rightarrow 4f^{3}$, $4f^{2} 6g \rightarrow 4f^{3}$,  
$4f^{2} 7g \rightarrow 4f^{3}$, and $4f^{2} 8g \rightarrow 4f^{3}$ transitions. 
The previous investigation  showed that strong electron-impact excitations 
occur from the ground configuration  to the $4f^{2} 5g$ and $4f^{2} 6g$ 
configurations \cite{2014jqsrt_136_108_alkauskas}. Populations of these 
configurations are not affected by the cascade emission process because the 
$4f^{2} 5g$, $4f^{2} 6g$, $4f^{2} 7g$, and $4f^{2} 8g$ configurations are highly 
excited ones. The cascade emission is responsible for increase of the spectral 
line intensities at 5 nm.

\begin{figure}
 \includegraphics[scale=0.5]{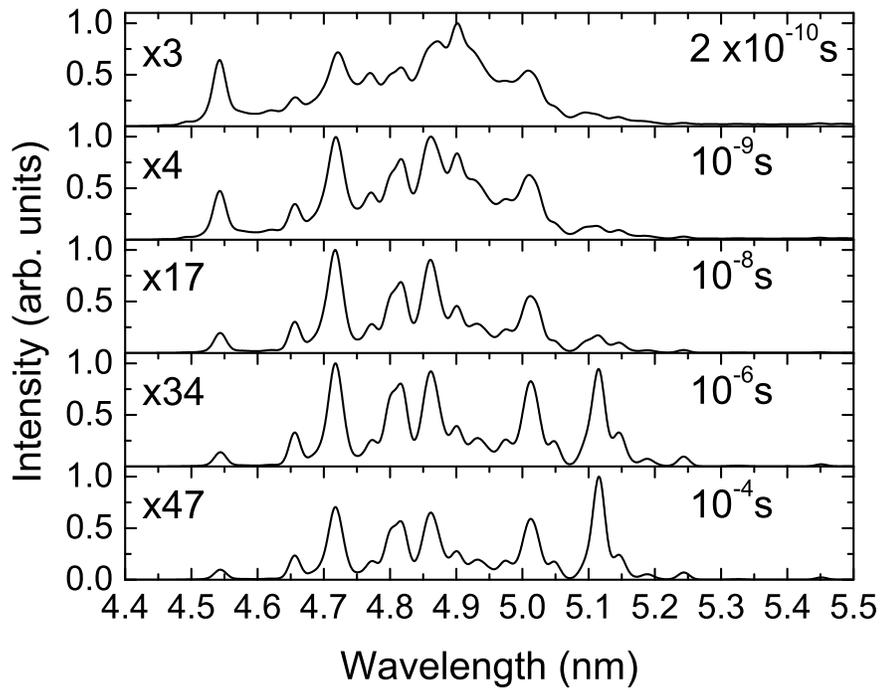}
 \caption{\label{fig4} Time-integrated spectra of cascade emission in the 
$4.4 - 5.5$ nm region. Times spent by ions outside the elecron beam are 
shown. The factor shows an increase of the line intensities compared to the 
CRM spectrum.} 
\end{figure}

Other interesting result of modeling is formation of lines in the $4.5 - 5.3$ nm 
region (Fig. \ref{fig3}). It seems that the structure of these lines is not so 
significantly affected by the cascade emission as in the $13 - 18$ nm range 
because the emission lines overlap in the CRM and cascade emission spectra. 
However, it can be seen from Fig. \ref{fig3} that the cascade emission spectrum 
is more structured than the CRM one and distribution of the line intensities is 
different. The CRM calculations show that the lines in the $4.5 - 5.3$ nm region 
correspond to the $4d^{9} 4f^{4} \rightarrow 4f^{3}$ and 
$4f^{2} 5d \rightarrow 4f^{3}$  transitions.  Nevertheless, other levels are 
involved in the line formation for the cascade emission spectrum compared to the 
CRM calculations. Figure \ref{fig4} shows how  line intensities in the region 
changes with time. It can be seen that many strong lines disappear from the 
spectrum while the other line intensities are significantly increased. 
The strongest lines in the CRM spectrum correspond to the 
$4d^{9} 4f^{4} \rightarrow 4f^{3}$ transitions (Table \ref{t2}) while the 
$4f^{2} 5d \rightarrow 4f^{3}$  transitions  dominate in the cascade emission 
spectrum (Table \ref{t3}). One can see that the distribution of the line 
intensities in the CRM calculations is more smooth compared with the cascade 
emission data. There are only a few strong lines in the cascade emission 
spectrum.


\begingroup 

\renewcommand{\arraystretch}{1.2}
\renewcommand{\tabcolsep}{2mm}
\scriptsize

\begin{flushleft}

\LTcapwidth 13cm

\begin{longtable}{ l l l l l l l l}

\caption{\label{t2}  The strongest lines of the CRM spectrum for $W^{25+}$ in 
the $4-7$ nm wavelength range. Wavelengths $\lambda$, relative intensities $I$, 
and indexes of initial $i$ and final $f$ levels are presented. 
$J$ stands for the total angular momentum quantum number.} \\

\hline
$\lambda$ (nm) & $I$ & $i$ & $f$ & $J_{i}$ & $J_{f}$ & Initial level & Final level \\
\hline
\endfirsthead
\caption[]{ (continued) }  \\
\hline
$\lambda$ (nm) & $I$ & $i$ & $f$ & $J_{i}$ & $J_{f}$ & Initial level & Final level \\
\hline
\endhead
\hline \multicolumn{4}{r}{\textit{Continued on next page}} \\
\endfoot
\hline
\endlastfoot

$4.546$ & $100$ & $1221$ & $31$ & $15/2$ & $17/2$ & $4d_{3/2}^{3}$  $(3/2)$  $4f_{5/2}^{2}$  $(2)$ $3/2$  $4f_{7/2}^{2}$  $(6)$  & $4f_{5/2}^{1}$   $ 4f_{7/2}^{2}$  $(6)$ \\ 
$4.656$ & $99$ & $1200$ & $31$ & $17/2$ & $17/2$ & $4d_{3/2}^{3}$  $(3/2)$  $4f_{5/2}^{2}$  $(4)$ $5/2$ $4f_{7/2}^{2}$  $(6)$  & $4f_{5/2}^{1}$   $4f_{7/2}^{2}$  $(6)$\\
$4.543$ & $91$ & $1186$ & $6$ & $13/2$ & $15/2$ & $4d_{3/2}^{3}$  $(3/2)$  $4f_{5/2}^{1}$  $(1)$  $4f_{7/2}^{3}$  $(15/2)$  & $4f_{5/2}^{1}$    $4f_{7/2}^{2}$  $(6)$\\
$4.544$ & $91$ & $1176$ & $3$ & $11/2$ & $13/2$ & $4d_{3/2}^{3}$  $(3/2)$  $4f_{5/2}^{2}$ $(4)$ $5/2$  $4f_{7/2}^{2}$  $(6)$  & $4f_{5/2}^{1}$    $4f_{7/2}^{2}$  $(6)$\\
$4.536$ & $87$ & $1204$ & $13$ & $11/2$ & $13/2$ & $4d_{3/2}^{3}$  $(3/2)$  $4f_{5/2}^{3}$  $(9/2)$ $3$  $4f_{7/2}^{1}$    & $4f_{5/2}^{2}$  $(4)$  $4f_{7/2}^{1}$  \\
$4.539$ & $84$ & $1220$ & $27$ & $11/2$ & $13/2$ & $4d_{3/2}^{3}$  $(3/2)$  $4f_{5/2}^{2}$  $(4)$ $5/2$  $4f_{7/2}^{2}$  $(4)$  & $4f_{5/2}^{1}$    $4f_{7/2}^{2}$  $(4)$\\
$4.543$ & $82$ & $1209$ & $18$ & $13/2$ & $15/2$ & $4d_{3/2}^{3}$  $(3/2)$  $4f_{5/2}^{1}$   $(1)$  $4f_{7/2}^{3}$  $(15/2)$ & $4f_{5/2}^{2}$  $(4)$  $4f_{7/2}^{1}$  \\
$4.544$ & $79$ & $1219$ & $29$ & $13/2$ & $15/2$ & $4d_{3/2}^{3}$  $(3/2)$  $4f_{5/2}^{1}$   $(1)$  $4f_{7/2}^{3}$  $(15/2)$ & $4f_{5/2}^{1}$    $4f_{7/2}^{2}$  $(6)$\\
$4.727$ & $78$ & $1137$ & $13$ & $13/2$ & $13/2$ & $4d_{3/2}^{3}$  $(3/2)$   $4f_{5/2}^{3}$  $(9/2)$ $4$  $4f_{7/2}^{1}$   & $4f_{5/2}^{2}$  $(4)$  $4f_{7/2}^{1}$  \\
$4.682$ & $75$ & $1188$ & $29$ & $15/2$ & $15/2$ & $4d_{3/2}^{3}$  $(3/2)$  $4f_{5/2}^{1}$   $(1)$  $4f_{7/2}^{3}$  $(15/2)$ & $4f_{5/2}^{1}$    $4f_{7/2}^{2}$  $(6)$\\
$4.701$ & $72$ & $1160$ & $18$ & $15/2$ & $15/2$ & $4d_{3/2}^{3}$  $(3/2)$  $4f_{5/2}^{3}$  $(9/2)$ $4$  $4f_{7/2}^{1}$  & $4f_{5/2}^{2}$  $(4)$  $4f_{7/2}^{1}$  \\
$4.721$ & $68$ & $1174$ & $27$ & $13/2$ & $13/2$ & $4d_{3/2}^{3}$  $(3/2)$  $4f_{5/2}^{2}$  $(4)$ $5/2$  $4f_{7/2}^{2}$  $(4)$ & $4f_{5/2}^{1}$    $4f_{7/2}^{2}$  $(4)$\\
$4.539$ & $66$ & $1167$ & $1$ & $9/2$ & $11/2$ & $4d_{3/2}^{3}$  $(3/2)$  $4f_{5/2}^{3}$  $(9/2)$ $3$  $4f_{7/2}^{1}$   & $4f_{5/2}^{2}$  $(4)$ $4f_{7/2}^{1}$  \\
$4.733$ & $65$ & $1121$ & $6$ & $15/2$ & $15/2$ & $4f_{7/2}^{2}$  $(6)$ $5d_{3/2}^{1}$   & $4f_{5/2}^{1}$    $4f_{7/2}^{2}$  $(6)$\\
$4.738$ & $59$ & $1183$ & $34$ & $11/2$ & $11/2$ & $4d_{3/2}^{3}$  $(3/2)$  $4f_{5/2}^{1}$   $(1)$  $4f_{7/2}^{3}$  $(11/2)$ & $4f_{7/2}^{3}$  $(11/2)$\\ 
$4.727$ & $55$ & $1095$ & $3$ & $13/2$ & $13/2$ & $4f_{5/2}^{1}$    $4f_{7/2}^{1}$   $4$  $5d_{3/2}^{1}$   & $4f_{5/2}^{1}$    $4f_{7/2}^{2}$  $(6)$\\
$4.572$ & $51$ & $1203$ & $12$ & $9/2$ & $11/2$ & $4d_{3/2}^{3}$  $(3/2)$  $4f_{5/2}^{2}$  $(4)$ $5/2$  $4f_{7/2}^{2}$  $(6)$ & $4f_{5/2}^{1}$    $4f_{7/2}^{2}$  $(6)$\\
$4.524$ & $49$ & $1223$ & $34$ & $9/2$ & $11/2$ & $4d_{3/2}^{3}$  $(3/2)$  $4f_{5/2}^{1}$   $(1)$  $4f_{7/2}^{3}$  $(11/2)$ & $4f_{7/2}^{3}$  $(11/2)$\\
$4.751$ & $47$ & $1054$ & $1$ & $11/2$ & $11/2$ & $4d_{3/2}^{3}$  $(3/2)$  $4f_{5/2}^{3}$  $(9/2)$ $(4)$  $4f_{7/2}^{1}$   & $4f_{5/2}^{2}$  $(4)$  $4f_{7/2}^{1}$  \\
$5.116$ & $44$ & $989$ & $31$ & $19/2$ & $17/2$ & $4d_{3/2}^{3}$  $(3/2)$  $4f_{5/2}^{2}$  $(4)$ $(7/2)$  $4f_{7/2}^{2}$  $(6)$ & $4f_{5/2}^{1}$    $4f_{7/2}^{2}$  $(6)$\\
\end{longtable}

\end{flushleft}

\endgroup

The spectral feature of lower intensity is visible at about 5.5 nm to 6 nm in 
the CRM spectrum but it is not seen in the cascade emission calculations. This 
additional lower intensity peak is presented in the fusion spectra 
\cite{2005jpb_38_3071_putterich} but it disappears from the EBIT spectra 
\cite{Radtke2001PhysRevA_64_012720}. The obtained results illustrate importance 
of the cascade emission of ions outside the electron beam in the EBIT device. 
To the best of our knowledge, these differences in the fusion and EBIT spectra  
have not been explained before. It has to be noted that group of lines in 
$5.5-6.0$ nm region seen in the CRM spectrum disappears from the cascade 
emission spectrum after about $10^{-8}$ s. The relative intensities of the 
lines decrease about two times after $2 \cdot 10^{-10}$ s and four times 
after $10^{-9}$ s compared to the intensity of the strongest line in the 
spectrum.  Since this group of lines is not seen in the EBIT spectrum it implies 
that the ions spend outside the beam in average more than $10^{-9}$ s. 


\begingroup 

\renewcommand{\arraystretch}{1.2}
\renewcommand{\tabcolsep}{2mm}
\scriptsize

\begin{flushleft}

\LTcapwidth 13cm

\begin{longtable}{ l l l l l l l l  }

\caption{\label{t3}  The strongest lines of the cascade emission spectrum for  
$W^{25+}$ in the $4-7$ nm wavelength range. Wavelengths $\lambda$, relative 
intensities $I$, and indexes of initial $i$ and final $f$ levels are presented. 
$J$ stands for the total angular momentum quantum number.} \\

\hline
$\lambda$ (nm) & $I$ & $i$ & $f$ & $J_{i}$ & $J_{f}$ & Initial level & Final level \\
\hline
\endfirsthead
\caption[]{ (continued) }  \\
\hline
$\lambda$ (nm) & $I$ & $i$ & $f$ & $J_{i}$ & $J_{f}$ & Initial level & Final level \\
\hline
\endhead
\hline \multicolumn{4}{r}{\textit{Continued on next page}} \\
\endfoot
\hline
\endlastfoot

$5.116$	 & $100$	 & $989$	 & $31$  & $19/2$	& $17/2$&	$4d_{3/2}^{3}$  $(3/2)$  $4f_{5/2}^{2}$  $(4)$  $7/2$  $4f_{7/2}^{2}$  $(6)$	&	$4f_{5/2}^{1}$    $4f_{7/2}^{2}$  $(6)$\\
$4.716$	 & $62$	 & $1121$	 & $6$ 	 &	$15/2$	&	$15/2$	&	$4f_{7/2}^{2}$  $(6)$  $5d_{5/2}^{1}$  	&	$4f_{5/2}^{1}$    $4f_{7/2}^{2}$  $(6)$\\
$4.818$	 & $52$	 & $1061$	 & $6$ 	 &	$15/2$	&	$15/2$	&	$4f_{5/2}^{1}$    $4f_{7/2}^{1}$   $5$  $5d_{5/2}^{1}$  	&	$4f_{5/2}^{1}$    $4f_{7/2}^{2}$  $(6)$\\
$4.800$	 & $43$	 & $1152$	 & $31$  &	$19/2$	&	$17/2$	&	$4f_{5/2}^{1}$    $4f_{7/2}^{1}$   $6$  $5d_{5/2}^{1}$  	&	$4f_{5/2}^{1}$    $4f_{7/2}^{2}$  $(6)$\\
$4.864$	 & $37$	 & $1029$	 & $6$ 	 &	$15/2$	&	$15/2$	&	$4f_{7/2}^{2}$  $(6)$  $5d_{3/2}^{1}$  	&	$4f_{5/2}^{1}$    $4f_{7/2}^{2}$  $(6)$\\
$5.007$	 & $29$	 & $1038$	 & $27$  &	$15/2$	&	$13/2$	&	$4f_{7/2}^{2}$  $(6)$  $5d_{3/2}^{1}$  	&	$4f_{5/2}^{1}$    $4f_{7/2}^{2}$  $(4)$\\
$4.856$	 & $26$	 & $1087$	 & $18$  &	$15/2$	&	$15/2$	&	$4f_{7/2}^{2}$  $(6)$  $5d_{5/2}^{1}$  	&	$4f_{5/2}^{2}$  $(4)$  $4f_{7/2}^{1}$  \\
$4.656$	 & $25$	 & $1200$	 & $31$  &	$17/2$	&	$17/2$	&	$4d_{3/2}^{3}$  $(3/2)$  $4f_{5/2}^{2}$  $(4)$ $5/2$  $4f_{7/2}^{2}$  $(6)$	&	$4f_{5/2}^{1}$    $4f_{7/2}^{2}$  $(6)$\\ 
$4.727$	 & $23$	 & $1095$	 & $3$ 	 &	$13/2$	&	$13/2$	&	$4f_{5/2}^{1}$    $4f_{7/2}^{1}$   $4$  $5d_{5/2}^{1}$  	&	$4f_{5/2}^{1}$    $4f_{7/2}^{2}$  $(6)$\\
$5.147$	 & $23$	 & $899$	 & $6$ 	 &	$17/2$	&	$15/2$	&	$4d_{3/2}^{3}$  $(3/2)$  $4f_{5/2}^{3}$  $(9/2)$ $5$  $4f_{7/2}^{1}$  	&	$4f_{5/2}^{1}$    $4f_{7/2}^{2}$  $(6)$\\
$5.014$	 & $22$	 & $956$	 & $6$ 	 &	$17/2$	&	$15/2$	&	$4d_{3/2}^{3}$  $(3/2)$  $4f_{5/2}^{2}$  $(4)$ $7/2$  $4f_{7/2}^{2}$  $(6)$	&	$4f_{5/2}^{1}$    $4f_{7/2}^{2}$  $(6)$\\
$5.024$	 & $21$	 & $1029$	 & $27$  &	$15/2$	&	$13/2$	&	$4f_{7/2}^{2}$  $(6)$  $5d_{3/2}^{1}$  	&	$4f_{5/2}^{1}$    $4f_{7/2}^{2}$  $(4)$\\
$5.106$	 & $18$	 & $956$	 & $18$  &	$17/2$	&	$15/2$	&	$4d_{3/2}^{3}$  $(3/2)$  $4f_{5/2}^{2}$  $(4)$ $7/2$  $4f_{7/2}^{2}$  $(6)$	&	$4f_{5/2}^{2}$  $(4)$  $4f_{7/2}^{1}$  \\
$5.048$	 & $17$	 & $1018$	 & $29$  &	$17/2$	&	$17/2$	&	$4d_{3/2}^{3}$  $(3/2)$ $4f_{5/2}^{1}$   $1$ $4f_{7/2}^{3}$  $(15/2)$	&	$4f_{5/2}^{1}$    $4f_{7/2}^{2}$  $(6)$\\
$4.901$	 & $16$	 & $1109$	 & $31$  &	$15/2$	&	$17/2$	&	$4f_{7/2}^{2}$  $(6)$  $5d_{3/2}^{1}$  	&	$4f_{5/2}^{1}$    $4f_{7/2}^{2}$  $(6)$\\
$4.701$	 & $15$	 & $1160$	 & $18$  &	$15/2$	&	$15/2$	&	$4d_{3/2}^{3}$  $(3/2)$  $4f_{5/2}^{3}$  $(9/2)$ $4$  $4f_{7/2}^{1}$  	&	$4f_{5/2}^{2}$  $(4)$  $4f_{7/2}^{1}$  \\
$4.773$	 & $13$	 & $1087$	 & $6$ 	 &	$15/2$	&	$15/2$	&	$4f_{7/2}^{2}$  $(6)$  $5d_{5/2}^{1}$  	&	$4f_{5/2}^{1}$    $4f_{7/2}^{2}$  $(6)$\\
$5.094$	 & $10$	 & $949$	 & $13$  &	$15/2$	&	$13/2$	&	$4d_{3/2}^{3}$  $(3/2)$  $4f_{5/2}^{1}$  $(4)$ $7/2$  $4f_{7/2}^{2}$  $(4)$	&	$4f_{5/2}^{2}$  $(4)$  $4f_{7/2}^{1}$ \\
$4.975$	 & $8$	 & $1061$	 & $27$  &	$15/2$	&	$13/2$	&	$4f_{5/2}^{1}$   $4f_{7/2}^{1}$   $5$  $5d_{5/2}^{1}$  	&	$4f_{5/2}^{1}$    $4f_{7/2}^{2}$  $(4)$\\
$5.244$	 & $8$	 & $899$	 & $18$  &	$17/2$	&	$15/2$	&	$4d_{3/2}^{3}$  $(3/2)$  $4f_{5/2}^{3}$  $(9/2)$ $5$  $4f_{7/2}^{1}$  	&	$4f_{5/2}^{2}$  $(4)$  $4f_{7/2}^{1}$  \\
\end{longtable}

\end{flushleft}
\endgroup

In addition, we have estimated influence of charge exchange process on formation 
of spectral lines due to interaction with neutrals. The captured electron 
occupies a state with principal quantum number 
$n_{c} \approx Z_{eff}^{3/4}$ ($Z_{eff}$ is the effective charge of the ion) 
\cite{2000prl_85_5090_beiersdorfer}. For the W$^{26+}$ ion, one can derive 
$n_{c} \approx 12$. The angular momentum of the captured electron is defined by 
the ion charge and the relative collision velocity $v$ (in atomic units): 
$l = (5 Z_{eff})^{0.5} v$ \cite{2000prl_85_5090_beiersdorfer}; that leads to 
$l=0$ for the considered collision energy. Again, the cascade emission from the 
$4f^{2} 12s$ configuration gives a large number of the lines in the $13 - 18$ 
nm range. It indicates that the charge exchange process is not important for the 
formation of the spectral lines from the W$^{25+}$ ion in the EBIT plasma. 
The same result was obtained for the W$^{13+}$ ion 
\cite{2013jqsrt_127_64_jonauskas} and for the higher ionization stages of the 
tungsten ions \cite{2006pra_74_042514_Ralchenko, 2007jpb_40_3861_ralchenko}.

\section{Conclusions}

The CRM with ensuing cascade emission have been studied for the W$^{25+}$ ion. 
It is demonstrated that the cascade emission is responsible for formation of 
some lines in the EBIT spectra of the W$^{25+}$ ion. These lines correspond to 
transitions among the levels with high $J$ values.

The relative intensity of lines at 5 nm is strongly increased in the cascade 
emission spectrum compared to the lines at the shorter wavelength side which 
are not affected by the cascade emission. The cascade emission produces only 
few strong lines in the region while the CRM calculations give more smooth 
distribution for the line intensities. The strongest lines in the CRM spectrum 
correspond to the $4d^{9} 4f^{4} \rightarrow 4f^{3}$ transitions while many 
lines from the $4f^{2} 5d \rightarrow 4f^{3}$  transitions appear in the cascade 
emission calculations. 

The CRM gives a spectrum with a complex structure of lines in the $13 - 18$ nm 
region contradicting the observations as well as cascade emission spectrum. 
Calculations show that the lines belong to the $4f^{2}5d \rightarrow 4f^{2}5p$ 
and $4f^{2}5p \rightarrow 4f^{2}5s$ transitions. 

The less intense line structure observed in fusion spectra at about 6 nm is 
reproduced by our CRM calculations. The missing structure of the lines in the 
EBIT measurements is explained by the cascade emission of ions outside the 
electron beam. The reason of the difference between the fusion and EBIT spectra 
for this wavelength region has never been determined before. Time-integrated 
study of the line intensities gives that the ions spend in average more than 
$10^{-9}$ s outside the electron beam.

Finally, our results demonstrate that the cascade emission has to be taken into 
account for the ions in intermediate ionization stages when the spectra from 
the  EBIT plasma are analyzed. The CRM alone does not provide a reasonable 
agreement with the measurements because it omits physical processes which occur 
after the ions leave the electron beam region.

\section*{Akcnowledgement}
This research was funded by European Social Fund under the Global Grant Measure 
(No.: VP1-3.1-\v{S}MM-07-K-02-015).


\begin{thebibliography}{27}%
\makeatletter
\providecommand \@ifxundefined [1]{%
 \@ifx{#1\undefined}
}%
\providecommand \@ifnum [1]{%
 \ifnum #1\expandafter \@firstoftwo
 \else \expandafter \@secondoftwo
 \fi
}%
\providecommand \@ifx [1]{%
 \ifx #1\expandafter \@firstoftwo
 \else \expandafter \@secondoftwo
 \fi
}%
\providecommand \natexlab [1]{#1}%
\providecommand \enquote  [1]{``#1''}%
\providecommand \bibnamefont  [1]{#1}%
\providecommand \bibfnamefont [1]{#1}%
\providecommand \citenamefont [1]{#1}%
\providecommand \href@noop [0]{\@secondoftwo}%
\providecommand \href [0]{\begingroup \@sanitize@url \@href}%
\providecommand \@href[1]{\@@startlink{#1}\@@href}%
\providecommand \@@href[1]{\endgroup#1\@@endlink}%
\providecommand \@sanitize@url [0]{\catcode `\\12\catcode `\$12\catcode
  `\&12\catcode `\#12\catcode `\^12\catcode `\_12\catcode `\%12\relax}%
\providecommand \@@startlink[1]{}%
\providecommand \@@endlink[0]{}%
\providecommand \url  [0]{\begingroup\@sanitize@url \@url }%
\providecommand \@url [1]{\endgroup\@href {#1}{\urlprefix }}%
\providecommand \urlprefix  [0]{URL }%
\providecommand \Eprint [0]{\href }%
\providecommand \doibase [0]{http://dx.doi.org/}%
\providecommand \selectlanguage [0]{\@gobble}%
\providecommand \bibinfo  [0]{\@secondoftwo}%
\providecommand \bibfield  [0]{\@secondoftwo}%
\providecommand \translation [1]{[#1]}%
\providecommand \BibitemOpen [0]{}%
\providecommand \bibitemStop [0]{}%
\providecommand \bibitemNoStop [0]{.\EOS\space}%
\providecommand \EOS [0]{\spacefactor3000\relax}%
\providecommand \BibitemShut  [1]{\csname bibitem#1\endcsname}%
\let\auto@bib@innerbib\@empty
\bibitem [{\citenamefont {{Bolt}}\ \emph {et~al.}(2002)\citenamefont {{Bolt}},
  \citenamefont {{Barabash}}, \citenamefont {{Federici}}, \citenamefont
  {{Linke}}, \citenamefont {{Loarte}}, \citenamefont {{Roth}},\ and\
  \citenamefont {{Sato}}}]{Bolt_2002jnm_307_43}%
  \BibitemOpen
  \bibfield  {author} {\bibinfo {author} {\bibnamefont
  {{Bolt}} \bibfnamefont {H}}, \bibinfo {author} {\bibnamefont {{Barabash}} \bibfnamefont {V}},
  \bibinfo {author} {\bibnamefont {{Federici}} \bibfnamefont {G}}, \bibinfo
  {author} {\bibnamefont {{Linke}} \bibfnamefont {J}}, \bibinfo {author}
  {\bibnamefont {{Loarte}} \bibfnamefont {A}}, \bibinfo {author}
  {\bibnamefont {{Roth}} \bibfnamefont {J}}, \ and\ \bibinfo {author}
  {\bibnamefont {{Sato}} \bibfnamefont {K}}.\ }\href@noop {}  {\bibfield
  {title} {\bibinfo  {title} {Plasma facing and high heat flux materials - needs for ITER and beyond.}\ }} {\bibfield
  {journal} {\bibinfo  {journal} {J Nucl Mater}\ } \bibinfo {year} {2002}; {\bibinfo {volume}
  {307}}:\ \bibinfo {pages} {43-52}}\BibitemShut
  {NoStop}%
\bibitem [{\citenamefont {P\"{u}tterich}\ \emph {et~al.}(2005)\citenamefont
  {P\"{u}tterich}, \citenamefont {Neu}, \citenamefont {Biedermann},
  \citenamefont {Radtke},\ and\ \citenamefont
  {Team}}]{2005jpb_38_3071_putterich}%
  \BibitemOpen
  \bibfield  {author} {\bibinfo {author} {\bibnamefont
  {P\"{u}tterich} \bibfnamefont {T}}, \bibinfo {author} {\bibnamefont {Neu} \bibfnamefont {R}},
  \bibinfo {author} {\bibnamefont {Biedermann} \bibfnamefont {C}}, \bibinfo
  {author} {\bibnamefont {Radtke} \bibfnamefont {R}} \ and\ \bibinfo {author}
  {\bibnamefont {ASDEX Upgrade Team}. }\ }
  {\bibinfo{title} {Disentangling the emissions of highly ionized tungsten in the range 4 - 14 nm.}}
  { {\bibinfo  {journal} {J Phys B: At Mol Opt Phys}\ } \bibinfo {year} {2005}; {\bibinfo {volume} {38}}:\
  \bibinfo {pages} {3071}}\BibitemShut {NoStop}%
\bibitem [{\citenamefont {Radtke}\ \emph {et~al.}(2007)\citenamefont {Radtke},
  \citenamefont {Biedermann}, \citenamefont {Fussmann}, \citenamefont {Schwob},
  \citenamefont {Mandelbaum},\ and\ \citenamefont
  {Doron}}]{2007apmidf_13_45_radtke}%
  \BibitemOpen
  \bibfield  {author} {\bibinfo {author} {\bibnamefont
  {Radtke} \bibfnamefont {R}}, \bibinfo {author} {\bibnamefont {Biedermann} \bibfnamefont {C}},
  \bibinfo {author} {\bibnamefont {Fussmann} \bibfnamefont {G}}, \bibinfo
  {author} {\bibnamefont {Schwob} \bibfnamefont {J}}, \bibinfo {author}
  {\bibnamefont {Mandelbaum} \bibfnamefont {P}} \ and\ \bibinfo {author}
  {\bibnamefont {Doron} \bibfnamefont {R.}}\ }\href@noop {} {\bibfield
  {title} {\bibinfo  {title} {Measured line spectra and calculated atomic physics data for highly charged tungsten ions.}\ }} {\bibfield
  {journal} {\bibinfo  {journal} {At  Plas Mater Interac Data
  Fusion}\ }\bibinfo {year} {2007;} {\bibinfo {volume} {13:}}\ \bibinfo {pages} {45}}\BibitemShut {NoStop}%
\bibitem [{\citenamefont {P\"{u}tterich}\ \emph {et~al.}(2008)\citenamefont
  {P\"{u}tterich}, \citenamefont {Neu}, \citenamefont {Dux}, \citenamefont
  {Whiteford},\ and\ \citenamefont {O'Mullane}}]{2008ppcp_50_085016_Putterich}%
  \BibitemOpen
   \bibfield  {author} {\bibinfo {author} {\bibnamefont
  {P\"{u}tterich}} \bibfnamefont {T}, \bibinfo {author} {\bibnamefont {Neu}} \bibfnamefont {R},
  \bibinfo {author} {\bibnamefont {Dux} \bibfnamefont {R}}, \bibinfo {author}
  { \bibnamefont {Whiteford} \bibfnamefont {a~D}} \ and\ \bibinfo {author}
  {\ \bibnamefont {O'Mullane} \bibfnamefont {M~G}}.\ } {\bibfield  {title} {\bibinfo  {title}
  {Modelling of measured tungsten spectra from ASDEX Upgrade and predictions for ITER.}\ }}{\bibfield  {journal} {\bibinfo  {journal}
  {Plas Phys Contr Fusion}\ }\bibinfo {year} {2008;} {\bibinfo {volume} {50}}:\
  \bibinfo {pages} {085016}}\BibitemShut {NoStop}%
\bibitem [{200(2009)}]{2009adas_biedermann}%
  \BibitemOpen
  \bibfield  {author} {\bibinfo {author} {\bibnamefont
  {Biedermann} \bibfnamefont {C}}.\ }
   {\bibfield  {title} {\bibinfo  {title} {Spectroscopy of highly charge ions with EBIT.}\ }} {\bibfield  {journal} {\bibinfo  {journal} {ADAS Workshop
  2009}}} \url{http://www.adas.ac.uk/talks2009.php} \BibitemShut {NoStop}%
\bibitem [{\citenamefont {Suzuki}\ \emph {et~al.}(2011)\citenamefont {Suzuki},
  \citenamefont {Harte}, \citenamefont {Kilbane}, \citenamefont {Kato},
  \citenamefont {Sakaue}, \citenamefont {Murakami}, \citenamefont {Kato},
  \citenamefont {Sato}, \citenamefont {Tamura}, \citenamefont {Sudo},
  \citenamefont {Goto}, \citenamefont {D'Arcy}, \citenamefont {Sokell},\ and\
  \citenamefont {O'Sullivan}}]{2011jpb_44_175004_suzuki}%
  \BibitemOpen
  \bibfield  {author} {\bibinfo {author} {\bibnamefont
  {Suzuki} \bibfnamefont {C}}, \bibinfo {author} {\ \bibnamefont {Harte} \bibfnamefont {C~S}},
  \bibinfo {author} {\bibnamefont {Kilbane} \bibfnamefont {D}}, \bibinfo
  {author} {\bibnamefont {Kato} \bibfnamefont {T}}, \bibinfo {author}
  {\ \bibnamefont {Sakaue} \bibfnamefont {H~A}}, \bibinfo {author}
  {\bibnamefont {Murakami} \bibfnamefont {I}}, \bibinfo {author}
  {\bibnamefont {Kato} \bibfnamefont {D}}, \bibinfo {author} {\bibnamefont {Sato} \bibfnamefont
  {K}}, \bibinfo {author} {\bibnamefont
  {Tamura} \bibfnamefont {N}}, \bibinfo {author} {\bibnamefont {Sudo} \bibfnamefont {S}},
  \bibinfo {author} {\bibnamefont {Goto} \bibfnamefont {M}}, \bibinfo {author}
  {\bibnamefont {D'Arcy} \bibfnamefont {R}}, \bibinfo {author} {\bibnamefont {Sokell} \bibfnamefont
  {E}} \ and\ \bibinfo {author} {\bibnamefont {O'Sullivan} \bibfnamefont
  {G.}}\ } {\bibfield  {title}
  {\bibinfo  {title} {Interpretation of spectral emission in the 20 nm region from tungsten ions observed in fusion device plasmas.}\ }} {\bibfield  {journal}
  {\bibinfo  {journal} {J Phys B: At Mol Opt  Phys}\ } {\bibinfo {year} {2011}};\ \bibinfo {volume} {44(17):}
  \bibinfo {pages} {175004}}\BibitemShut {NoStop}%
\bibitem [{\citenamefont {P\"{u}tterich}\ \emph {et~al.}(2013)\citenamefont
  {P\"{u}tterich}, \citenamefont {Jonauskas}, \citenamefont {Neu},
  \citenamefont {Dux},\ and\ \citenamefont
  {Team}}]{2013acp_1545_132_putterich}%
  \BibitemOpen
  \bibfield  {author} {\bibinfo {author} {\bibnamefont
  {P\"{u}tterich} \bibfnamefont {T}}, \bibinfo {author} {\bibnamefont
  {Jonauskas} \bibfnamefont {V}}, \bibinfo {author} {\bibnamefont {Neu} \bibfnamefont {R}},
  \bibinfo {author} {\bibnamefont {Dux} \bibfnamefont {R}} \ and\ \bibinfo
  {author} {\bibfnamefont {ASDEX Upgrade}\ \bibnamefont {Team}}.\ } {\bibfield  {title} {\bibinfo
  {title} {The extreme ultraviolet emissions of W$^{23+}$($4f^5$) .}\ }}{\bibfield  {journal} {\bibinfo
  {journal} {AIP Conf Proc}\ } {\bibinfo {year} {2013}};\ \bibinfo
  {volume} {1545(1):} \bibinfo {pages} {132-142}}\BibitemShut {NoStop}%
\bibitem [{\citenamefont {Alkauskas}\ \emph {et~al.}(2014)\citenamefont
  {Alkauskas}, \citenamefont {Rynkun}, \citenamefont {Gaigalas}, \citenamefont
  {KynienÃÂÃÂ}, \citenamefont {Kisielius}, \citenamefont {KuÃÂÃÂas},
  \citenamefont {Masys}, \citenamefont {Merkelis},\ and\ \citenamefont
  {Jonauskas}}]{2014jqsrt_136_108_alkauskas}%
  \BibitemOpen
  \bibfield  {author} {\bibinfo {author} \bibnamefont
  {Alkauskas} {\bibfnamefont {A}}, \bibinfo {author} {\bibnamefont {Rynkun} \bibfnamefont {P}},
  \bibinfo {author} {\bibnamefont {Gaigalas} \bibfnamefont {G}}, \bibinfo
  {author} {\bibnamefont {Kynien\.{e}} \bibfnamefont {A}}, \bibinfo {author}
  {\bibnamefont {Kisielius} \bibfnamefont {R}}, \bibinfo {author}
  {\bibnamefont {Ku\v{c}as} \bibfnamefont {S}}, \bibinfo {author}
  {\bibnamefont {Masys} \bibfnamefont {\v{S}}}, \bibinfo {author} {\bibnamefont {Merkelis} \bibfnamefont
  {G}} \ and\ \bibinfo {author} {\bibnamefont {Jonauskas} \bibfnamefont
  {V.}}\ } {\bibfield  {title}
  {\bibinfo  {title} {Theoretical investigation of spectroscopic properties of W$^{25+}$.}\ }}{\bibfield  {journal}
  {\bibinfo  {journal} {J Quant Spectrosc Radiat Transfer}\ }
  {\bibinfo {year} {2014}};\ \bibinfo {volume} {136:} \bibinfo {pages}
  {108-118}}\BibitemShut {NoStop}%
\bibitem [{201(2013)}]{2013jqsrt_127_64_jonauskas}%
  \BibitemOpen
  \bibfield  {author} {\bibinfo {author} \bibnamefont
  {Jonauskas} {\bibfnamefont {V}},  \bibinfo {author} {\bibnamefont {Masys} \bibfnamefont {\v{S}}},
   \bibinfo {author} {\bibnamefont {Kynien\.{e}} \bibfnamefont {A}},  \bibinfo
  {author} {\bibnamefont {Gaigalas} \bibfnamefont {G}}. \ } {\bibfield  {title} {\bibinfo
  {title} {Cascade emission in electron beam ion trap plasma.}\ }} {\bibfield  {journal} {\bibinfo
  {journal} {J Quant Spectrosc Radiat Transfer}\ } {\bibinfo
  {year} {2013}};\ \bibinfo {volume} {127:} \bibinfo {pages}
  {64-69}}\BibitemShut {NoStop}%
\bibitem [{\citenamefont {Jonauskas}\ \emph {et~al.}(2003)\citenamefont
  {Jonauskas}, \citenamefont {Partanen}, \citenamefont {Ku\v{c}as},
  \citenamefont {Karazija}, \citenamefont {Huttula}, \citenamefont {Aksela},\
  and\ \citenamefont {Aksela}}]{2003jpb_36_4403_jonauskas}%
  \BibitemOpen
  \bibfield  {author} {\bibinfo {author} {\bibnamefont
  {Jonauskas} \bibfnamefont {V}}, \bibinfo {author} {\bibnamefont {Partanen} \bibfnamefont {L}},
  \bibinfo {author} {\bibnamefont {Ku\v{c}as} \bibfnamefont {S}}, \bibinfo
  {author} {\bibnamefont {Karazija} \bibfnamefont {R}}, \bibinfo {author}
  {\bibnamefont {Huttula} \bibfnamefont {M}}, \bibinfo {author} {\bibnamefont {Aksela} \bibfnamefont
  {S}} \ and\ \bibinfo {author} {\bibnamefont {Aksela} \bibfnamefont
  {H.}}\ }
  {\bibfield  {title} {\bibinfo  {title} {Auger cascade satellites following $3d$ ionization in xenon.}\ }} {\bibfield  {journal} {\bibinfo  {journal} {J Phys B:  At 
  Mol Opt Phys}\ }{\bibinfo
  {year} {2003}};\ \bibinfo {volume} {36(22):} \bibinfo {pages} {4403-4416}}\BibitemShut {NoStop}%
\bibitem [{\citenamefont {Jonauskas}\ \emph {et~al.}(2008)\citenamefont
  {Jonauskas}, \citenamefont {Karazija},\ and\ \citenamefont
  {Ku\v{c}as}}]{2008jpb_41_215005_jonauskas}%
  \BibitemOpen
  \bibfield  {author} {\bibinfo {author} {\bibnamefont
  {Jonauskas} \bibfnamefont {V}}, \bibinfo {author} {\bibnamefont {Karazija} \bibfnamefont {R}},
  \ and\ \bibinfo {author} {\bibnamefont {Ku\v{c}as} \bibfnamefont {S.}} \
  } {\bibfield  {title}
  {\bibinfo  {title} {The essential role of many-electron Auger transitions in the cascades following the photoionization of $3p$ and $3d$ shells of Kr.}\ }} {\bibfield  {journal}
  {\bibinfo  {journal} {J Phys B: At Mol Opt Phys}\ } {\bibinfo {year} {2008}};\ \bibinfo {volume} {41(21):}
  \bibinfo {pages} {215005(5pp)}}\BibitemShut {NoStop}%
\bibitem [{\citenamefont {Palaudoux}\ \emph {et~al.}(2010)\citenamefont
  {Palaudoux}, \citenamefont {Lablanquie}, \citenamefont {Andric},
  \citenamefont {Ito}, \citenamefont {Shigemasa}, \citenamefont {Eland},
  \citenamefont {Jonauskas}, \citenamefont {Ku\v{c}as}, \citenamefont
  {Karazija},\ and\ \citenamefont {Penent}}]{2010pra_82_043419_palaudoux}%
  \BibitemOpen
  \bibfield  {author} {\bibinfo {author} {\bibnamefont
  {Palaudoux} \bibfnamefont {J}}, \bibinfo {author} {\bibnamefont
  {Lablanquie} \bibfnamefont {P}}, \bibinfo {author} {\bibnamefont {Andric} \bibfnamefont {L}},
  \bibinfo {author} {\bibnamefont {Ito} \bibfnamefont {K}}, \bibinfo {author}
  {\bibnamefont {Shigemasa} \bibfnamefont {E}}, \bibinfo {author}
  {\ \bibnamefont {Eland} \bibfnamefont {J~H~D}}, \bibinfo {author}
  {\bibnamefont {Jonauskas} \bibfnamefont {V}}, \bibinfo {author}
  {\bibnamefont {Ku\v{c}as} \bibfnamefont {S}}, \bibinfo {author}
  {\bibnamefont {Karazija} \bibfnamefont {R}} \ and\ \bibinfo {author}
  {\bibnamefont {Penent} \bibfnamefont {F.}}\ } {\bibfield  {title} {\bibinfo  {title}
  {{Multielectron spectroscopy: Auger decays of the krypton $3d$ hole.}}\ }} {\bibfield  {journal} {\bibinfo  {journal}
  {{Phys Rev A}}\ } {\bibinfo {year} {{2010}}};\ \bibinfo {volume}
  {{82(4):}} \bibinfo {pages} {{043419}}}\BibitemShut {NoStop}%
\bibitem [{\citenamefont {Jonauskas}\ \emph {et~al.}(2011)\citenamefont
  {Jonauskas}, \citenamefont {Ku\ifmmode~\check{c}\else \v{c}\fi{}as},\ and\
  \citenamefont {Karazija}}]{2011pra_84_053415_jonauskas}%
  \BibitemOpen
  \bibfield  {author} {\bibinfo {author} {\bibnamefont
  {Jonauskas} \bibfnamefont {V}}, \bibinfo {author} {\bibnamefont
  {Ku\ifmmode~\check{c}\else \v{c}\fi{}as} \bibfnamefont {S}} \ and\ \bibinfo {author}
  {\bibnamefont {Karazija} \bibfnamefont {R.}}\ } {\bibfield  {title} {\bibinfo  {title} {Auger decay of 3$p$-ionized krypton.}\ }} {\bibfield  {journal} {\bibinfo  {journal} {Phys
  Rev A}\ }{\bibinfo {year} {2011}};\ \bibinfo {volume} {84:}
  \bibinfo {pages} {053415}}\BibitemShut {NoStop}%
\bibitem [{\citenamefont {Gillaspy}\ \emph {et~al.}(1995)\citenamefont
  {Gillaspy}, \citenamefont {Aglitskiy}, \citenamefont {Bell}, \citenamefont
  {Brown}, \citenamefont {Chantler}, \citenamefont {Deslattes}, \citenamefont
  {Feldman}, \citenamefont {Hudson}, \citenamefont {Laming}, \citenamefont
  {Meyer}, \citenamefont {Morgan}, \citenamefont {Pikin}, \citenamefont
  {Roberts}, \citenamefont {Ratliff}, \citenamefont {Serpa}, \citenamefont
  {Sugar},\ and\ \citenamefont {Takacs}}]{1995ps_59_392_gillaspy}%
  \BibitemOpen
    \bibfield  {author} {\bibinfo {author} {\bibnamefont
  {Gillaspy} \bibfnamefont {J}}, \bibinfo {author} {\bibnamefont {Aglitskiy} \bibfnamefont {Y}},
  \bibinfo {author} {\bibnamefont {Bell} \bibfnamefont {E}}, \bibinfo {author}
  {\bibnamefont {Brown} \bibfnamefont {C}}, \bibinfo {author} {\bibnamefont {Chantler} \bibfnamefont
  {C}}, \bibinfo {author} {\bibnamefont {Deslattes}\bibfnamefont
  {R}}, \bibinfo {author} {\bibnamefont {Feldman} \bibfnamefont
  {U}}, \bibinfo {author} {\bibnamefont {Hudson} \bibfnamefont
  {L}}, \bibinfo {author} {\bibnamefont {Laming} \bibfnamefont
  {J}}, \bibinfo {author} {\bibnamefont {Meyer} \bibfnamefont
  {E}}, \bibinfo {author} {\bibnamefont {Morgan} \bibfnamefont
  {C}}, \bibinfo {author} {\bibnamefont {Pikin} \bibfnamefont
  {A}}, \bibinfo {author} {\bibnamefont {Roberts} \bibfnamefont
  {J}}, \bibinfo {author} {\bibnamefont {Ratliff} \bibfnamefont
  {L}}, \bibinfo {author} {\bibnamefont {Serpa} \bibfnamefont
  {F}}, \bibinfo {author} {\bibnamefont {Sugar} \bibfnamefont
  {J}} \ and\ \bibinfo {author} {\bibnamefont {Takacs} \bibfnamefont
  {E}}.\ } {\bibfield  {title} {\bibinfo  {title}
  {{Overview of the electron-beam ion-trap program at NIST.}}\ }}{\bibfield  {journal} {\bibinfo  {journal}
  {{Phys Scr}}\ } {\bibinfo {year} {{1995}}};\ \bibinfo {volume}
  {T59:} \bibinfo {pages} {{392-395}}}\BibitemShut {NoStop}%
\bibitem [{\citenamefont {Liang}\ \emph {et~al.}(2009)\citenamefont {Liang},
  \citenamefont {Lopez-Urrutia}, \citenamefont {Baumann}, \citenamefont {Epp},
  \citenamefont {Gonchar}, \citenamefont {Lapierre}, \citenamefont {Mokler},
  \citenamefont {Simon}, \citenamefont {Tawara}, \citenamefont {Maeckel},
  \citenamefont {Yao}, \citenamefont {Zhao}, \citenamefont {Zou},\ and\
  \citenamefont {Ullrich}}]{2009apj_702_838_liang}%
  \BibitemOpen
    \bibfield  {author} {\bibinfo {author} {\bibnamefont
  {Liang} \bibfnamefont {G~Y}}, \bibinfo {author} {\bibnamefont
  {Lopez-Urrutia} \bibfnamefont {J~R~C}}, \bibinfo {author} {\bibnamefont {Baumann} \bibfnamefont {T~M}}, \bibinfo {author} {\bibnamefont {Epp} \bibfnamefont {S~W}},
  \bibinfo {author} {\bibnamefont {Gonchar} \bibfnamefont {A}}, \bibinfo
  {author} {\bibnamefont {Lapierre} \bibfnamefont {A}}, \bibinfo {author}
  {\bibnamefont {Mokler} \bibfnamefont {P~H}}, \bibinfo {author}
  {\bibnamefont {Simon} \bibfnamefont {M~C}}, \bibinfo {author}
  {\bibnamefont {Tawara} \bibfnamefont {H}}, \bibinfo {author} {\bibnamefont {Maeckel} \bibfnamefont {V}}, \bibinfo {author} {\bibnamefont {Yao} \bibfnamefont {K}}, \bibinfo {author} {\bibnamefont
  {Zhao} \bibfnamefont {G}}, \bibinfo {author} {\bibnamefont {Zou} \bibfnamefont {Y}} \ and\
  \bibinfo {author} {\bibnamefont {Ullrich} \bibfnamefont {J}}.} {\bibfield  {title} {\bibinfo
  {title} {{Experimental investigations of ion charge distributions, effective electron densities, and electron-ion cloud overlap in electron beam ion trap plasma using extreme-ultraviolet spectroscopy.}}\ }} {\bibfield  {journal} {\bibinfo
  {journal} {{Astr J}}\ } {\bibinfo {year} {{2009}}};\
  \bibinfo {volume} {702(2):} \bibinfo {pages} {{838-850}}}\BibitemShut {NoStop}%
\bibitem [{\citenamefont {Chen}\ \emph {et~al.}(2004)\citenamefont {Chen},
  \citenamefont {Beiersdorfer}, \citenamefont {Heeter}, \citenamefont
  {Liedahl}, \citenamefont {Naranjo-Rivera}, \citenamefont {Tr\"{a}bert},
  \citenamefont {Gu},\ and\ \citenamefont {Lepson}}]{2004apj_611_598_chen}%
  \BibitemOpen
    \bibfield  {author} {\bibinfo {author} {\bibnamefont
  {Chen} \bibfnamefont {H}}, \bibinfo {author} {\bibnamefont {Beiersdorfer} \bibfnamefont {P}}, \bibinfo {author} {\bibnamefont {Heeter} \bibfnamefont {L~A}}, \bibinfo
  {author} {\bibnamefont {Liedahl} \bibfnamefont {D~A}}, \bibinfo {author}
  {\bibnamefont {Naranjo-Rivera} \bibfnamefont {K~L}}, \bibinfo {author}
  {\bibnamefont {Tr\"{a}bert} \bibfnamefont {E}}, \bibinfo {author}
  {\bibnamefont {Gu} \bibfnamefont {M~F}} \ and\ \bibinfo {author}
  {\bibnamefont {Lepson} \bibfnamefont {J~K}}.} {\bibfield  {title}
  {\bibinfo  {title} {Experimental and Theoretical Evaluation of Density-sensitive N VI, Ar XIV, and Fe XXII Line Ratios.}\ }} {\bibfield  {journal}
  {\bibinfo  {journal} {Astr J}\ } {\bibinfo {year}
  {2004;}}\ \bibinfo {volume} {611:} \bibinfo {pages} {598-604}}\BibitemShut  {NoStop}%
\bibitem [{\citenamefont {Jonauskas}\ \emph {et~al.}(2007)\citenamefont
  {Jonauskas}, \citenamefont {Ku\v{c}as},\ and\ \citenamefont
  {Karazija}}]{Jonauskas2007jpb_40_2179}%
  \BibitemOpen
   \bibfield  {author} {\bibinfo {author} {\bibnamefont {Jonauskas} \bibfnamefont {V}}, \bibinfo {author} {\bibnamefont {Ku\v{c}as} \bibfnamefont {S}} \ and\ \bibinfo {author} {\bibnamefont {Karazija} \bibfnamefont {R.}}\ } {\bibfield
  {title} {\bibinfo  {title} {On the interpretation of the intense emission of tungsten ions at about 5 nm.}\ }} {\bibfield
  {journal} {\bibinfo  {journal} {J Phys B: At Mol Opt Phys}\ }
  {\bibinfo {year} {2007;}}\ \bibinfo {volume} {40(11):} \bibinfo {pages} {2179-2188}}\BibitemShut {NoStop}%
\bibitem [{\citenamefont {Gu}(2008)}]{2008cjp_86_675_Gu}%
  \BibitemOpen
  \bibfield  {author} {\bibinfo {author} {\bibnamefont {Gu} \bibfnamefont {M~F.}\ }\ } {\bibfield  {title} {\bibinfo
  {title} {The flexible atomic code.}\ }} {\bibfield  {journal} {\bibinfo
  {journal} {Can J Phys}\ } {\bibinfo {year} {2008;}} \bibinfo {volume} {86:} \bibinfo {pages} {675-689}}\BibitemShut {NoStop}%
\bibitem [{\citenamefont {{Radtke}}\ \emph {et~al.}(2001)\citenamefont
  {{Radtke}}, \citenamefont {{Biedermann}}, \citenamefont {{Schwob}},
  \citenamefont {{Mandelbaum}},\ and\ \citenamefont
  {{Doron}}}]{Radtke2001PhysRevA_64_012720}%
  \BibitemOpen
  \bibfield  {author} {\bibinfo {author} {\bibnamefont {Radtke} \bibfnamefont {R}}, \bibinfo {author} {\bibnamefont {Biedermann} \bibfnamefont {C}}, \bibinfo {author} {\bibnamefont {Schwob} \bibfnamefont {J~L}}, \bibinfo {author} {\bibnamefont {Mandelbaum} \bibfnamefont {P}} \ and\ \bibinfo {author} {\bibnamefont {Doron} \bibfnamefont {R.}}\ } \bibfield
  {title} {\bibinfo  {title} {Line and band emission from tungsten ions with charge $21+$ to $45+$ in the $45 - 70$ {\AA} range.}} \bibfield {journal} {\bibinfo  {journal} {Phys Rev A} \bibinfo {year} {2001;}\ \bibinfo {volume} {64(1):} \bibinfo {pages} {012720}}\BibitemShut  {NoStop}%
\bibitem [{\citenamefont {Jonauskas}\ \emph {et~al.}(2010)\citenamefont
  {Jonauskas}, \citenamefont {Kisielius}, \citenamefont {{Kynien\. e}},
  \citenamefont {{S.~Ku{\v {c}}as}},\ and\ \citenamefont
  {Norrington}}]{2010pra_81_012506_jonauskas}%
  \BibitemOpen
  \bibfield  {author} {\bibinfo {author} {\bibnamefont
  {Jonauskas} \bibfnamefont {V}}, \bibinfo {author} {\bibnamefont
  {Kisielius} \bibfnamefont {R}}, \bibinfo {author} {\bibnamefont {{Kynien\.
  e}} \bibfnamefont {A}}, \bibinfo {author} {\bibnamefont {{Ku{\v {c}}as S}}} \ and\ \bibinfo
  {author} {\bibnamefont {Norrington} \bibfnamefont {P~H.}}\ } {\bibfield  {title} {\bibinfo  {title} {Magnetic dipole transitions in $4d^N$ configurations of tungsten ions.}\ }}  {\bibfield  {journal} {\bibinfo  {journal} {Phys Rev A}\ }{\bibinfo  {year} {2010}};\ \bibinfo {volume} {81:} \bibinfo {pages}
  {012506}}\BibitemShut {NoStop}%
\bibitem [{\citenamefont {Jonauskas}\ \emph {et~al.}(2012)\citenamefont
  {Jonauskas}, \citenamefont {Gaigalas},\ and\ \citenamefont
  {Ku\v{c}as}}]{2012adndt_98_19_jonauskas}%
  \BibitemOpen
   \bibfield  {author} {\bibinfo {author} {\bibnamefont {Jonauskas} \bibfnamefont {V}}, \bibinfo {author} {\bibnamefont {Gaigalas} \bibfnamefont {G}} \ and\ \bibinfo {author} {\bibnamefont {Ku\v{c}as} \bibfnamefont {S.}}\
  }{\bibfield  {title} {\bibinfo {title} {Relativistic calculations for M1-type transitions in configurations of W$^{29+}$ - W$^{37+}$ ions.}\ }} {\bibfield  {journal} {\bibinfo
  {journal} {At Data Nucl Data Tables}\ } {\bibinfo {year}
  {2012}};\ \bibinfo {volume} {98(1):} \bibinfo {pages} {19-42}} \BibitemShut
  {NoStop}%
\bibitem [{\citenamefont {Karazija}(1996)}]{Karazija1}%
  \BibitemOpen
  \bibfield  {author} {\bibinfo {author} {\bibnamefont
  {Karazija} \bibfnamefont {R.}}\ }{{\bibinfo {title} {Introduction to the
  Theory of X-Ray and Electronic Spectra of Free Atoms.}}}\ \bibinfo
  {address} {New York}:\ \bibinfo {publisher} {Plenum Press};\ \bibinfo {year}
  {1996}\BibitemShut {NoStop}%
\bibitem [{\citenamefont {{Ku{\v {c}}as}}\ \emph {et~al.}(1997)\citenamefont
  {{Ku{\v {c}}as}}, \citenamefont {Jonauskas},\ and\ \citenamefont
  {Karazija}}]{1997ps_55_667_kucas}%
  \BibitemOpen
  \bibfield  {author} {\bibinfo {author} {\bibnamefont
  {{Ku{\v {c}}as}} \bibfnamefont {S}}, \bibinfo {author} {\bibnamefont
  {Jonauskas} \bibfnamefont {V}} \ and\ \bibinfo {author} {\bibnamefont
  {Karazija} \bibfnamefont {R.}}\ } {\bibfield   {title} {\bibinfo  {title} {Global characteristics of atomic spectra and their use for the analysis of spectra. IV. Configuration interaction effects.}\ }}  {\bibfield   {journal} {\bibinfo  {journal} {Phys Scr}\ } {\bibinfo {year}
  {1997}};\ \bibinfo {volume} {55(6):} \bibinfo {pages} {667-675}}\BibitemShut
  {NoStop}%
\bibitem [{\citenamefont {Clementson}\ \emph {et~al.}(2010)\citenamefont
  {Clementson}, \citenamefont {Beiersdorfer}, \citenamefont {Magee},
  \citenamefont {McLean},\ and\ \citenamefont
  {Wood}}]{2010jpb_43_144009_clementson}%
  \BibitemOpen
  \bibfield  {author} {\bibinfo {author} {\bibnamefont  {Clementson} \bibfnamefont {J}}, \bibinfo {author} {\bibnamefont
  {Beiersdorfer} \bibfnamefont {P}}, \bibinfo {author} {\bibnamefont {Magee} \bibfnamefont {E~W}}, \bibinfo {author} {\bibnamefont {McLean} \bibfnamefont {H~S}} \
  and\ \bibinfo {author} {\bibnamefont {Wood} \bibfnamefont {R~D.}}\ } \bibfield  {title} {\bibinfo
  {title} {Tungsten spectroscopy relevant to the diagnostics of ITER divertor plasmas.}\ }  {\bibfield  {journal} {\bibinfo
  {journal} {J Phys B: At Mol Opt Phys}\
  } {\bibinfo {year} {2010}};\ \bibinfo {volume} {43(14):} \bibinfo
  {pages} {144009}}\BibitemShut {NoStop}%
\bibitem [{\citenamefont {Beiersdorfer}\ \emph {et~al.}(2000)\citenamefont
  {Beiersdorfer}, \citenamefont {Olson}, \citenamefont {Brown}, \citenamefont
  {Chen}, \citenamefont {Harris}, \citenamefont {Neill}, \citenamefont
  {Schweikhard}, \citenamefont {Utter},\ and\ \citenamefont
  {Widmann}}]{2000prl_85_5090_beiersdorfer}%
  \BibitemOpen
  \bibfield  {author} {\bibinfo {author} {\bibnamefont
  {Beiersdorfer} \bibfnamefont {P}}, \bibinfo {author} {\bibnamefont
  {Olson} \bibfnamefont {R~E}}, \bibinfo {author} {\bibnamefont {Brown} \bibfnamefont {G~V}},
  \bibinfo {author} {\bibnamefont {Chen} \bibfnamefont {H}}, \bibinfo {author}
  {\bibnamefont {Harris} \bibfnamefont {C~L}}, \bibinfo {author}
  {\bibnamefont {Neill} \bibfnamefont {P~A}}, \bibinfo {author}
  {\bibnamefont {Schweikhard} \bibfnamefont {L}}, \bibinfo {author}
  {\bibnamefont {Utter} \bibfnamefont {S~B}} \ and\ \bibinfo {author}
  {\bibnamefont {Widmann} \bibfnamefont {K.}}} \bibfield  {title} {\bibinfo  {title}
  {X-Ray Emission Following Low-Energy Charge Exchange Collisions of Highly Charged Ions.}} {\bibfield  {journal} {\bibinfo  {journal}
  {Phys Rev Lett}\ } {\bibinfo {year} {2000}};\ \bibinfo {volume}
  {85(24):} \bibinfo {pages} {5090-5093}}\BibitemShut {NoStop}%
\bibitem [{\citenamefont {Ralchenko}\ \emph {et~al.}(2006)\citenamefont
  {Ralchenko}, \citenamefont {Tan}, \citenamefont {Gillaspy}, \citenamefont
  {Pomeroy},\ and\ \citenamefont {Silver}}]{2006pra_74_042514_Ralchenko}%
  \BibitemOpen
  \bibfield  {author} {\bibinfo {author} {\bibnamefont
  {Ralchenko} \bibfnamefont {Y}}, \bibinfo {author} {\bibnamefont {Tan} \bibfnamefont {J~N}},
  \bibinfo {author} { \bibnamefont {Gillaspy} \bibfnamefont {J~D}}, \bibinfo
  {author} {\bibnamefont {Pomeroy} \bibfnamefont {J~M}} \ and\ \bibinfo
  {author} {\bibnamefont {Silver} \bibfnamefont {E.}}} \bibfield  {title} {\bibinfo  {title} {Accurate modeling of benchmark x-ray spectra from highly charged ions of tungsten.}\ } {\bibfield  {journal} {\bibinfo  {journal}
  {Phys Rev A}\ } {\bibinfo {year} {{2006}}};\ \bibinfo {volume}
  {74(4):} \bibinfo {pages} {{042514}}}\BibitemShut {NoStop}%
\bibitem [{\citenamefont {Ralchenko}\ \emph {et~al.}(2007)\citenamefont
  {Ralchenko}, \citenamefont {Reader}, \citenamefont {Pomeroy}, \citenamefont
  {Tan},\ and\ \citenamefont {Gillaspy}}]{2007jpb_40_3861_ralchenko}%
  \BibitemOpen
  \bibfield  {author} {\bibinfo {author} {\bibnamefont {Ralchenko} \bibfnamefont {Y}}, \bibinfo {author} {\bibnamefont {Reader} \bibfnamefont {J}}, \bibinfo {author} {\bibnamefont {Pomeroy} \bibfnamefont {J~M}\ }, \bibinfo {author} {\bibnamefont {Tan} \bibfnamefont {J~N}\ } \ and\ \bibinfo {author} {\bibnamefont {Gillaspy} \bibfnamefont {J~D.}\ }\ } {\bibfield  {title} {\bibinfo {title} {Spectra of W$^{39+}$-W$^{47+}$ in the 12-20 nm region observed with an EBIT light source.}\ }} {\bibfield  {journal} {\bibinfo {journal} {J Phys B: At Mol Opt Phys}\ } {\bibinfo {year} {2007}};\ \bibinfo {volume} {40(19):} \bibinfo {pages} {3861-3875}}\BibitemShut {NoStop}%
\end{thebibliography}
\end{document}